\documentstyle{xray-wuerzburg-95}

\input psfig.tex

\newcommand{\be}{\begin{equation}} 
\newcommand{\ee}{\end{equation}} 

\def\gsim{\lower 2pt \hbox{$\, \buildrel {\scriptstyle >}\over
         {\scriptstyle \sim}\,$}}
\def\lsim{\lower 2pt \hbox{$\, \buildrel {\scriptstyle <}\over
         {\scriptstyle \sim}\,$}}

\newcommand{\Msun}{\mbox{$M_{\odot}\;$}}

\begin{document}
 
\title{Surface Temperature of Magnetized Neutron Stars}
\author{Dany Page \and A. Sarmiento}
\institute{Instituto de Astronom\'{\i}a, UNAM, M\'{e}xico D.F., M\'{e}xico;
           page@astroscu.unam.mx \& ansar@astroscu.unam.mx
           \vspace{-12.mm}}
\maketitle

\begin{figure}
\vspace{-4.5cm}
\psfig{figure=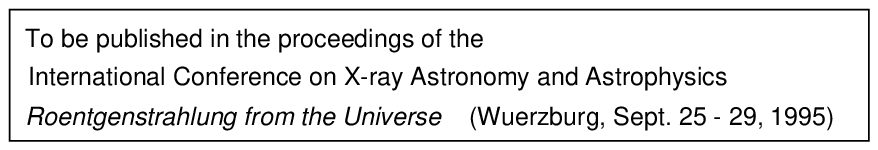}
\vspace{3.4cm}
\end{figure}
\vspace{-1.4cm}

\begin{abstract}
We show that the expected inhomogeneous temperature distribution induced at
the surface of a neutron star by the anisotropy of heat transport in
the magnetized envelope allows us to understand quite well the observed
pulse profiles of the four nearby pulsars for which surface thermal emission
has been detected.
However, due to gravitational lensing, dipolar magnetic fields are not
adequate and the observed high pulsed fractions force us to include a
quadrupolar component.
\vspace{-7.mm}
\end{abstract}

\vspace{-6.mm}
\section{INTRODUCTION}
\vspace{-2.mm}

The definite detections of surface thermal emission from four nearby
neutron stars by {\em ROSAT} (\"Ogelman 1995) opened up a new era in
the study of these objects: we are at the long last {\em seeing} an
isolated neutron star (Page 1995b).
All four objects: PSR 0833-45 (Vela; \"Ogelman, Finley, Zimmermann
1993), PSR 0656+14 (Finley, \"Ogelman, Kizilo\u{g}lu 1992), PSR
0630+178 (Geminga; Halpern, Holt 1992) and PSR 1055-52 (\"Ogelman \&
Finley 1993), show a two component spectrum with surface thermal
emission in the soft band and a hard tail.
Most important is that the soft thermal component is pulsed
at the 10 -- 30\% level for all four stars.
The hard tail is pulsed except maybe in the case of Vela (\"Ogelman 1995).
The origin of the hard tails is not clear yet although thermal
emission from the hot polar caps seems to be presently getting the
preference over a power-law emission.

A natural interpretation of the observed pulsations, in the soft X-ray band,
 is that the surface
temperature is not uniform: if it is determined by the heat flow from the hot
interior through the envelope then the presence of a strong magnetic field
will {\em a fortiori} induce large temperature inhomogeneities
(see, e.g., Yakovlev \& Kaminker 1994).
For a given field structure,
the temperature at any point of the neutron star surface can then
be easily calculated (Page 1995a), as a function of the local field strength
and of its angle $\Theta_B$ with the normal, and the distribution of surface 
temperature can be generated.

\vspace{-4.mm}
\section{RESULTS}
\vspace{-2.mm}

\begin{figure}
\psfig{figure=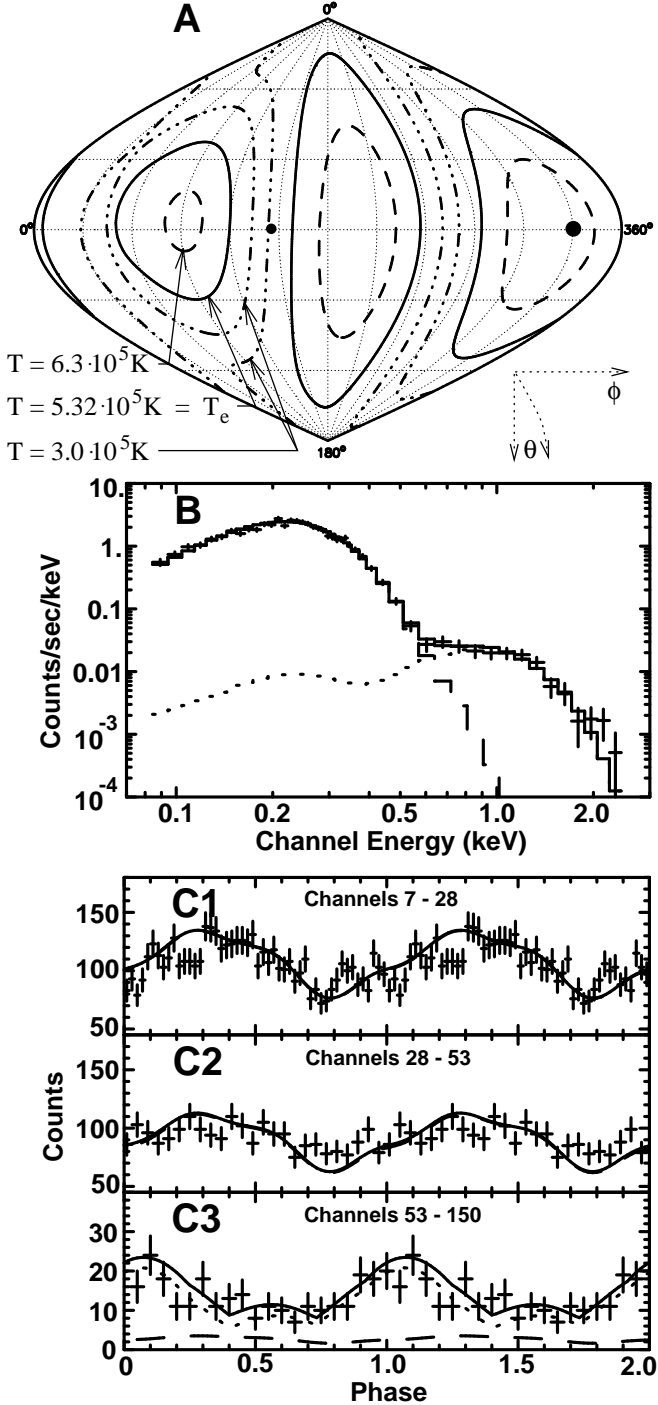}
\vspace{-5.mm}
\caption[]{A model for Geminga's soft X-ray emission (see text).
{\bf A)} Surface temperature distribution induced by a dipole+quadrupole field.
(Equal area mapping with $\theta$ vertically and $\phi$ horizontally).
{\bf B)} Spectral fit (surface thermal emission + hot polar caps).
{\bf C)} Fit of the pulse profiles in three channel bands.
The rotation axis is along $\theta = 0$ and the observer is at 90$^\circ$
as well as the dipole.
In B \& C: main surface emission in dashed, polar caps emission in dotted
and total emission in continuous.
C1 and C2 show that the polar caps make almost negligible contribution to the
low energy range and dominate the high energy range (as is also obvious from
B)}
\end{figure}

Magnetic field effects in the atmosphere are also extremely important
but as a first step we neglect them and restrict ourselves to consider only
blackbody (BB) emission.
Inclusion of these effects is in process and results will be reported later.

An important role is played by gravity through lensing: the observable
amplitude of the pulsations is very strongly reduced and if we assume
a simple dipolar field geometry, with reasonable neutron star masses
and radii, the resulting amplitudes are smaller than what is observed
(Page 1995a).
It is doubtful that inclusion of atmospheric effects will significantly
alter the conclusion:
{\em dipolar fields are totally insufficient to explain the observations}
(Page 1995a, b).
Moreover, the predicted shapes of the light curves are also very
distinct from what is observed.
A similar conclusion has been reached by Possenti {\em et al.} (1995) who 
performed a more detailed analysis of PSR 0656+14 along the same lines.

There are of course no strong {\em a priori} reasons for the surface field
to be dipolar and the previous conclusion is not really surprising.
The next obvious step is the inclusion of a quadrupolar component.
In general, this component will induce a very complicated surface temperature
distribution and will actually even reduce the amplitude of the pulsations.
However, if the orientation of the quadrupole with respect to the dipole is
adequately chosen then it is possible to increase substantially the
pulsations, even with very strong lensing
(e.g., 1.4 \Msun star with 8 km radius).
{\em With dipole+quadrupole magnetic fields it is possible to reproduce both
the spectrum and the pulse profile of the four pulsars we study}
(Page \& Sarmiento 1996).
The needed strength of the quadrupole is slightly lower than the dipole one:
if much lower, the quadrupole has no effect but if larger it will dominate
over the dipole and produce several distinct warm regions (where $\Theta_B
\sim 0^\circ$), flattening thus the light curves.
Possible evidence for the presence of a significant quadrupole has
already been brought up by high frequency radio observations of several
pulsars (Kuz'min 1992).
The four pulsars with detected surface thermal emission have, to our
knowledge, unfortunately not been observed at high radio
frequencies and it is thus not possible to compare the quadrupole we
need with radio observations.

We show in Fig. 1 an example of surface temperature distribution induced by a
dipole+quadrupole field and the resulting fit of the observed spectrum and 
light-curves of Geminga:
$M$ = 1.4 \Msun star, $R$ = 12 km ($R^\infty$ = 14.82 km) at a distance
$D$ = 185 pc and an effective temperature $T_e^\infty$ = 4.3 $\times 10^5$ K.
Two hot polar caps are added to fit the hard tail.
Both caps have a temperature $T^\infty$ of 2.7 $\times 10^6$ K and diameters of
0.55 and 0.40 degree: 
the positions of the polar caps (dots on panel A) have been chosen to fit
the light curve in channel band 53 -- 150.
Pulsations of the observed amplitude can be reproduced despite of
gravitational lensing in this model of 1.4 \Msun star with a radius of
12 km; at smaller radii fits are still possible but very restrictive
on
the quadrupole component.
The data are from Halpern \& Ruderman (1993).

An important feature of BB emission is that the pulsed fraction
always {\em increases} with photon energy in the energy range where
surface emission is detected, no matter the surface temperature distribution
(Page \& Sarmiento 1996).
This feature does not correspond to what is observed in the case of Geminga:
in Fig~1 the decrease in pulse amplitude in channel band 28~--~53 (C2) with
respect to the band 7~--~28 (C1) is not reproduced by this
model a may require a very different model (Page, Shibanov \& Zavlin 1995).

\vspace{-4.mm}
\section{CONCLUSIONS}
\vspace{-2.mm}

Reasonable surface magnetic field configurations (dipole + quadrupole)
allow us to interpret the observed pulse profiles as to due surface temperature
inhomogeneities induced by anisotropy of heat transport in the neutron star
magnetized envelope.
We used only BB spectra but the energy dependence of the observed pulsed
fraction in Geminga is already an indication that BB spectra are not adequate.
Another indication is the BB fit of the Vela spectrum which requires a 
distance of $\sim 1,500$ pc instead of the $\sim 500$ pc pulsar distance
(\"Ogelman {\em et al.} 1993) while magnetized hydrogen atmosphere spectra
are quite successful (Page, Shibanov \& Zavlin 1996).
A complete study will definitely need the inclusion of magnetic effects
both in the envelope and the atmosphere but our conclusion about the 
need of a quadrupolar component will certainly not be altered by the
use of realistic atmosphere models.

 
\end {document}